\documentclass[twocolumn,showpacs,amsmath,amssymb,prd,nofootinbib]{revtex4}
\usepackage{epsfig}
\def\sss{\scriptscriptstyle}
\def\^#1{^{\sss #1}}
\def\_#1{_{\sss #1}}
\def\beq{\begin{equation}}
\def\eeqno#1{\label{#1}\end{equation}}

\def\kms{{\rm ~km/s}}
\def\cmss{{\rm cm/s^{2}}}
\def\kpc{~{\rm Kpc}}

\def\msun{M\_{\odot}}
\def\az{a\_{0}}

\def\l0{\ell\_{0}}

\def\s{\sigma}
\def\l{\lambda}

\def\a{\alpha}

\def\vinf{V\_\infty}

\def\vmax{V_{max}}
\def\rM{r\_M}

\begin{document}
\title{Fast-rotating galaxies do not depart from the MOND mass-asymptotic-speed relation}
\author{Mordehai Milgrom}
\affiliation{Department of Particle Physics and Astrophysics, Weizmann Institute}

\begin{abstract}
Ogle et al. (2019) have fallaciously argued recently that fast-rotating disc galaxies break with the predictions of MOND.
Specifically, the six fastest rotators of the twenty three galaxies in their sample appear to have higher rotational speeds than is consistent with the MOND relation between the baryonic mass of a galaxy, $M$, and its `rotational speed', $V$. Ogle et al. further describe this claimed departure as a sharp break in the observed $M-V$  relation above some speed, from a logarithmic slope near the MOND-predicted value of $4$, to a much shallower slope of $\approx 0$.
These claims are, however, artefacts of plotting apples on a plot of oranges.
Ogle et al. use in their analysis the {\it maximal} rotational speed of the galaxies, $\vmax$, not the {\it asymptotic} one, $\vinf$, which appears in the MOND prediction, $\vinf^4=MG\az$ ($\az$ is the MOND acceleration constant), and which was used in previous analyses testing MOND. Plotting their $M$ vs. $\vmax$ pairs on a plot of $M$ vs. $\vinf$ from Lelli et al. (2016), they arrive erroneously at the above tension with MOND.  Now, the optical ($H_\a$) rotation curves used by Ogle et al. are far too short reaching to probe the asymptotic regime, and determine $\vinf$ (most of their $\vmax$ values  occur at radii that are merely 1-2 disc scale lengths).
However, it is well documented for fast rotators with observed, extended, HI rotation curves, that they can have $\vmax$ that is considerably larger than the MOND-relevant $\vinf$ -- e.g. Noordermeer and Verheijen (NV) in papers from 2007. For example, the highest-velocity disc in the sample of NV has $\vmax\approx 490\kms$, but $\vinf\approx 250\kms$. NV also show explicitly that in a (MOND-irrelevant) $M-\vmax$ plot, the high-speed galaxies fall off the power-law line defined by the lower-speed ones, creating a break in the $M$ vs. $V$ relation, in just the way claimed by Ogle et al. But, when plotting the MOND-relevant $M$ vs. $\vinf$, they find for the same galaxies, that all fall near the same power law, without a break.
\end{abstract}
\maketitle

\section{\label{introduction} Introduction}
One of the salient predictions of MOND \cite{milgrom83} is a relation between the total baryonic mass, $M$, of an isolated galaxy,\footnote{Or any other mass for that matter} and the asymptotic rotational speed, $\vinf$, of test particle on circular orbits around the galaxy. This is the MOND mass-asymptotic-speed relation (MASR).\footnote{Another MOND prediction is that the rotational speed is asymptotically constant; so its value is well defined.}
\beq MG\az=\vinf^4,  \eeqno{masr}
where $\az\approx 1.2\times 10^{-8}\cmss$ is the MOND acceleration constant. (For reviews of MOND see Refs. \cite{fm12,milgrom14c}.)
\par
Prediction (\ref{masr}) has been tested many times and vindicated (see, e.g., Refs. \cite{sanders96,nv07,mcgaugh11,lelli16,li18,lelli19}, and further references given in the last three).
\par
Luminosity-velocity, or mass-velocity plots for disc galaxies have been drawn many time in the literature, starting with the seminal photometric-magnitude-velocity plots of Tully and Fisher, known as Tully-Fisher relations. A class of such plots is called `baryonic Tully-Fisher relation' (BTFR) \cite{mcgaugh00}.
\par
By `baryonic' in BTFR, it is understood that one uses the total baryonic mass of the galaxy, not only some measure of the stellar mass (e.g., the luminosity), as was done in the original Tully-Fisher plot, and in many of its successors.
\par
The impetus to plot baryonic mass, $M$, not luminosities, is due to MOND, where it is $M$ that appears in the MOND prediction. This was first suggested if Ref. \cite{mb88} which analyzed the gas-rich dwarf DDO 154.
\par
However, calling a plot a BTFR does not specify the other entry; to wit, which velocity measure one uses: the HI width? the maximum speed, $\vmax$? some best-fit value, $V\_{flat}$, to the flat part of the rotation curve? the asymptotic speed, $\vinf$? Calling  different plots by the same name, as is found in the literature, can lead to misunderstandings and misrepresentations: a prime example of Bacon's {\it idola fori}.
\par
MOND dictates that we use $\vinf$. So , to avoid confusion, it is preferable to use the term MASR -- not the less concrete `BTFR' --  for a relation of the baryonic mass vs. the asymptotic velocity, or a close proxy for it.
\par
It has been demonstrated repeatedly that plotting $M$ vs. $\vinf$ minimizes the scatter, and yields a tighter observed relation (see, e.g., Refs. \cite{nv07,lelli19}).\footnote{One cannot really be sure that $\vinf$ has been reached and measured. Thus, those analyses use a velocity measure termed $V\_{flat}$, which is some best fit to the velocity in the `flat part' of the rotation curve. This $V\_{flat}$ may sometimes still be a little higher than $\vinf$. See the discussion in Sec. 4.2 of Ref. \cite{li18}.} This fact points to $\vinf$ as superior over other velocity measures, from the phenomenological point of view.
But, when discussing MOND, it is not a question of which tool is phenomenologically sharper. It is {\it imperative} that we use $\vinf$, not some other measure.
\par
For example, in low-surface-brightness galaxies, whose rotation curves increase slowly from the center out, the HI line velocity width -- sometimes used in `BTFR' analyses -- may be rather lower than $\vinf$, biasing the lower-end $M$ vs. $V$ slope to lower values than predicted by MOND.
\par
Using $\vmax$ when comparing with MOND is also wrong, because for high-mass, fast rotators,
$\vmax$ can be substantially higher than $\vinf$.
\par
The reason for this within the framework of MOND is that fast rotation correlates strongly with high surface density; high in the sense of MOND, namely with mean accelerations above $\az$. This means that such rotators are within their MOND radius $\rM\equiv(MG/\az)^{1/2}$.
This, in turn, means that their rotation curves tend to show a hump, with region of `Keplerian' decline within $\rM$ towards $\vinf$.
This effect is augmented by the fact that fast-rotator galaxies tend to have considerable, even dominant, bulges, which are even more concentrated than the discs, and have higher accelerations.
See, for example, Figs. 1-2 of Ref. \cite{milgrom83a} for the calculated MOND rotation curves for various model galaxies with different bulge-to-disc mass ratios, and different ratios of the galaxy size to its MOND radius.
\par
Ogle et al. \cite{ogle19} have recently measured and studied optical, $H_\a$ rotation curves of twenty three relatively-far, fast-rotating disc galaxies, with $\vmax=240-570\kms$. Plotting their $M-\vmax$ pairs on the `BTFR' of Ref. \cite{lelli16} -- which agrees very well with the MOND prediction (\ref{masr}) -- they find that the six highest $\vmax$ galaxies in their sample depart from the relation of Ref. \cite{lelli16}, lying to its high-velocity side.
From this Ogle et al.  conclude that these fast rotators disagree with the MOND prediction, and that the slope of the phenomenological `BTFR' breaks at $V\approx 350\kms$, from $\sim 4$ to $\sim 0$.
\par
Such fast-rotator disc galaxies (say, with  $\vmax\ge 250\kms$) are by no means a new discovery. Furthermore, the rotation curves of quite a few such fast rotators were analyzed successfully in MOND.  No notable, untoward behavior is seen in the earlier MOND analysis of these rotation curves.
\par
The record holder for many years had been  UGC 12591 with an $H_\a$ rotation curve given in Ref. \cite{giov86}, showing a maximum rotation speed of $\approx 500\kms$.
\par
Other examples are discussed in
Ref. \cite{sn07}, who analyzed in MOND a sample of seventeen disc galaxies with extended, HI rotation curves, using data from Ref. \cite{noord07}. Nine of the seventeen have $\vmax\ge 250\kms$ (six with $\vmax\ge 300\kms$); the highest being $\vmax=490\kms$, but with $\vinf\approx 250$. The second highest has $\vmax=390\kms$, with $\vinf\approx 320$.
Both these galaxies gave very good MOND fits with reasonable mass-to-light ratios. The analysis of Ref. \cite{sn07}, which, unlike Ref. \cite{ogle19}, separated the bulge and disc contributions, also show clearly how the presence of the bulges contributes materially to the appearance of declining rotation curves.
\par
Additional examples of such fast rotators can be found in the SPARC sample, analyzed in MOND in Ref. \cite{li18}, with the rotation curves themselves shown in Fig. 12 of Ref. \cite{milgrom20}.
\par
In Sec. \ref{ogle}, I explain why the results of Ogle et al. are based on fallacious reasoning, and
Sec. \ref{concl} raises additional potential worries regarding their analysis.

\section{Comments on the analysis of Ogle et al. \label{ogle}}
In short, the claims of Ogle et al. are fallacious if only because they use $\vmax$ and not $\vinf$ (see Sec. \ref{concl} for additional potential issues).
Atypically for analyses of this kind, Ogle et al. do not show their measured rotation curves, except for one (the highest-$\vmax$) case; so I have no information on their shape. But in any event, it is clear from the information given that
the radii they explore are much too small to have reached the asymptotic regime, and measure $\vinf$.
\par
Figure 2 of Ogle et al. plots their $M$ vs. $\vmax$ pairs superimposed on a plot of $M$ vs. $V\_{flat}$ (approximating $\vinf$), with the best-fit power law for the latter, from  Ref. \cite{lelli16} (which is consistent with the prediction of MOND). It shows that seventeen of the of the twenty three galaxies in the sample of Ogle et al. do fall near the best-fit line -- within the scatter that is seen at all speeds.
The six highest-$\vmax$ galaxies appear as outliers.\footnote{Notably, these are also, by and large, the farthest in the sample (at distances of $D=90-135\kpc$).}
The highest-$\vmax$ point lies a factor $\approx 2$ off in velocity.
The second highest $\vmax$ point is off by $\approx 35$ percent in velocity. However, the error quoted on its deduced $\vmax$ is large ($\pm 90\kms$); so it lies off the line by just $\approx 1.5\s$. Furthermore, according to the text of Ogle et al., the rotation curve of this galaxy is still rising at the last measured point, which is at a radius that is {\it less than their quoted scale length for the disc}. Clearly neither $\vmax$ and certainly not $\vinf$ have been measured for this galaxy. The other four velocities are $\sim 30$ percent off the line.
\par
From Table 1 of Ref. \cite{ogle19} we see that their deduced maximal speeds occur at only a few disc scale length: For five of the six rogues it occurs at 1-2 scale lengths, and for one at 3 scale length. From both MOND theory and observations of nearer-by, fast rotators, this is much too short a stretch for reaching $\vinf$.
\par
We thus have to draw on what we know about the nearer-by fast rotators with extended rotation curves: For slower rotators ($\vmax\lesssim 200\kms$), we have typically, but not universally, $\vmax\approx\vinf$ (the maximum is reached on the asymptotic part). However, for faster rotators, we have generically $\vmax>\vinf$, typically by tens of percents.
This can clearly be seen, for example, in Fig. 7 of Ref. \cite{noord07}, or in Fig. 4 of Ref. \cite{lelli19}, who plot the ratio $\vmax/\vinf$ against different galaxy characteristics.
\par
The occurrence of higher $\vmax/\vinf$ in fast rotators could explain the conclusions of Ogle et al. as an artifact.
For example, one of the seven fast rotators in the seventeen-galaxy sample studied in MOND by Sanders and Noordermeer \cite{sn07} (the one with highest speed) has $\vmax\approx 490\kms$, and $\vmax/\vinf\approx 2$. The next two fastest ($\vmax\approx 390\kms$, and $315\kms$) have $\vmax/\vinf\approx 1.2$.
\par
Furthermore, the six rouge galaxies in Ogle et al. all have even higher velocities, with $\vmax>435\kms$ in all six, and we expect them, according to the above trend, to have typically even higher values of $\vmax/\vinf$.
\par
Exactly this state of affairs was already described in 2007 in Ref. \cite{nv07} entitled `The high mass end of the Tully Fisher relation'.
To quote directly from the abstract of Ref. \cite{nv07}  (underscored by me): ``Using a combination of K-band photometry and {\it high-quality rotation curves,} we show that in traditional formulations of the TF relation (using the width of the global HI profile {\it or the maximum rotation velocity}), galaxies with rotation velocities larger than $200\kms$ lie systematically to the right of the relation defined by less massive systems, {\it causing a characteristic `kink' in the relations...}
We also show that many of {\it the galaxies with the largest offsets have declining rotation curves and that the change in slope largely disappears when we use the asymptotic rotation velocity as kinematic parameter}. {\it The remaining deviations from linearity can be removed when we simultaneously use the total baryonic mass} (stars + gas) instead of the optical or near-infrared luminosity.'' (See also Fig. 6 of Ref. \cite{nv07}.)
\par
As explained in Sec. \ref{introduction}, the `kink' in the $M-\vmax$ plot is due to transition from $\vmax\approx\vinf$ at lower speeds to $\vmax>\vinf$ at higher speeds.
\par
Similar appearance of declining rotation curves in fast rotators is described in Ref. \cite{genzel17} for high-redshift galaxies, and is discussed in Ref. \cite{milgrom17} in light of MOND.
\par
Another case of a misuse of $\vmax$ is in Ref. \cite{dai12}, who plot (their Fig. 9) $(M,~\vmax)$ for UGC 12591 on the relation found in Ref. \cite{mcgaugh05}, which correctly uses $V\_{flat}$ as a representative of $\vinf$ [and which agrees exactly with the MOND relation (\ref{masr})]. Reference \cite{dai12} estimate the galaxy's baryonic mass at $M\sim 10^{12}\msun$, and claim that this galaxy lies off the MOND relation.  However, one can see from the rotation curve (Fig. 3 of Ref. \cite{giov86}) that it declines from its $\vmax\approx 500\kms$ at about $10\kpc$ (corrected for a more updated Hubble constant) to about $420\kms$ at the last measured radius of $\approx 28\kpc$, but is still decreasing there. This rotation curve is also from $H_\a$ velocities and does not extend far enough to reach $\vinf$.
For the above baryonic mass we would expect $\vinf\approx 355\kms$.

\section{Further potential issues\label{concl}}
Unfortunately, except for one example, Ref. \cite{ogle19} do not show their rotation curves. This makes it difficult to assess the possible presence of further systematics, beyond the misuse of $\vmax$ instead of $\vinf$.
\par
By and large, these $H_\a$ rotation curve should be rather inferior to the high-quality HI rotation curves that have been used to test MOND in the past. The former are not only short reaching compared with the latter's extended curves; they are also based on velocity along the `major axis' of the galaxy, compared with HI rotation curves that are based on full two-dimensional velocity maps, which enable one not only to measure the rotation curve to much larger extents, but also afford one control over inclination changes due to warps, for example.
\par
The one rotation curve Ogle et al. do present -- that for their highest $\vmax$ -- is very {\it atypical} of fast rotators with high-quality measured rotation curve. It rises slowly in the inner parts instead of jumping quickly to a high value and then declining, as is known to be the case for fast rotators with HI rotation curves. This galaxy is, however, very nearly edge-on -- with a quoted inclination of $81$ degrees (two other of the six rogue galaxies also have high inclinations of 79 and 85 degrees, respectively). It might be then that the shape of the $H_\a$ rotation curve is highly distorted, at least in the inner parts, by extinction effects at the $H_\a$ wavelength. If extinction of the line is important, its measured wavelength does not represent the rotational speed at the line-of-sight tangent midpoint; so does not measure the required rotational speed.
 (Stacy McGaugh, private communication).
\par
Further distortions to the deduced rotation curves in edge-on galaxies, even if extinction is unimportant, may result from using the ($H_\a$) line center for determining the rotational velocity, as is done by Ogle et al.
(see e.g., Ref. \cite{allen79}, Federico Lelli, private communication).
\par
Such potential issues need further looking into, when more details of the analysis are made available.
\par
I thank Federico Lelli and Stacy McGaugh for useful comments.


\begin{thebibliography}{}
\bibitem{milgrom83}M. Milgrom (1983). A modification of the Newtonian dynamics as a possible alternative to the hidden mass hypothesis. Astrophys. J. 270, 365.
\bibitem{fm12}B. Famaey \& S. McGaugh (2012). Modified Newtonian Dynamics (MOND): Observational Phenomenology and Relativistic Extensions. Liv. Rev. Rel. 15, 10.
\bibitem{milgrom14c}M. Milgrom. (2014 - continually updated). The MOND paradigm of modified dynamics. Scholarpedia, 9(6), 31410.
\bibitem{sanders96}R. H.Sanders (1996). The Published Extended Rotation Curves of Spiral Galaxies: Confrontation with Modified Dynamics. Astrophys. J. 473, 117.
\bibitem{nv07}E. Noordermeer \& M.A.W. Verheijen (2007). The high-mass end of the Tully-Fisher relation.
    Mon. Not. Roy. Astron. Soc. 381, 1463.
\bibitem{mcgaugh11}S.S. McGaugh (2011). Novel Test of Modified Newtonian Dynamics with Gas Rich Galaxies. Phys. Rev. Lett. 106, 121303.
\bibitem{lelli16}F. Lelli \& al. (2016). The Small Scatter of the Baryonic Tully-Fisher Relation. Astrophys. J. Lett. 816, L14.
\bibitem{li18}Li, P., Lelli, F., McGaugh, S., \& Schombert, J. (2018). Fitting the radial acceleration relation to individual SPARC galaxies. Astron. Astrophys. 615, A3.
\bibitem{lelli19}F. Lelli \& al. (2019). The baryonic Tully-Fisher relation for different velocity definitions and implications for galaxy angular momentum. Mon. Not. Roy. Astron. Soc. 484, 3267.
\bibitem{mcgaugh00}S.S. McGaygh \& al. (2000). The Baryonic Tully-Fisher Relation. Astrophys. J. Lett. 533, L99.
\bibitem{mb88}M. Milgrom \& E. Braun (1988). The Rotation Curve of DDO 154: A Particularly Acute Test of the Modified Dynamics. Astrophys. J. 334, 130.
\bibitem{milgrom83a}M. Milgrom (1983). A modification of the Newtonian dynamics - Implications for galaxies.  Astrophys. J. 270, 371.

\bibitem{ogle19}P.M. Ogle \& al. (2019). A Break in Spiral Galaxy Scaling Relations at the Upper Limit of Galaxy Mass. Astrophys. J. Lett. 884, L11.

\bibitem{giov86}R. Giovanelli \& al. (1986). UGC 12591: The Most Rapidly Rotating Disk Galaxy.
Astrophy. J. Lett. 301, L7.
\bibitem{sn07}R.H. Sanders \& E. Noordermeer (2007). Confrontation of MOdified Newtonian Dynamics with the rotation curves of early-type disc galaxies. Mon. Not. Roy. Astron. Soc. 379, 702.
\bibitem{noord07}E. Noordermeer \& al. (2007). The mass distribution in early-type disc galaxies: declining rotation curves and correlations with optical properties. Mon. Not. Roy. Astron. Soc. 376, 1513.
\bibitem{milgrom20}M. Milgrom (2020). MOND vs. dark matter in light of historical parallels. Stud. Hist. Philos. Mod. Phys., in press. arXiv:1910.04368.
\bibitem{genzel17}R. Genzel \& al. (2017). Strongly baryon-dominated disk galaxies at the peak of galaxy formation ten billion years ago. Nature 543, 397G.
\bibitem{milgrom17}M. Milgrom (2017). High-redshift rotation curves and MOND. arXiv:1703.06110.
\bibitem{dai12}X. Dai \& al. (2012). XMM-Newton Detects a Hot Gaseous Halo in the Fastest Rotating Spiral Galaxy UGC 12591. Astrophys. J. 755, 107.
\bibitem{mcgaugh05}S.S. McGaugh (2005). The Baryonic Tully-Fisher Relation of Galaxies with Extended Rotation Curves and the Stellar Mass of Rotating Galaxies. Astrophys. J. 632, 859.
\bibitem{allen79}R.J. Allen \& R. Sancisi (1979). Neutral hydrogen observations of the edge-on disk galaxy NGC 891.
    Astron. Astrophys. 74, 73.
\end{thebibliography}
\end{document}